\documentclass{article}

\usepackage{tikz}
\usepackage[a4paper, portrait, margin=1in]{geometry}

\usepackage[round]{natbib}
\usepackage[colorlinks=true, citecolor=red, linkcolor=blue, urlcolor=green]{hyperref}
\usepackage{amsmath}
\usepackage{amsfonts}
\usepackage{cases}
\usepackage[affil-it]{authblk}

\usepackage{graphicx}
\graphicspath{{./images}}

\title{The Impact of Node Addition and Deletion on Network Production Fluctuations}

\author{Mahdi Kohan Sefidi}
\affil{University of Khatam}
\date{Jun. 2023}

\begin{document}
\maketitle
\begin{flushleft}

\end{flushleft}
\begin{abstract}
    Production networks, dynamic systems of firms linked through input-output relationships, transmit microeconomic shocks into macroeconomic fluctuations. While prior studies often assume static networks, real-world economies feature continuous firm entry (node addition) and exit (node deletion). We develop a probabilistic model to analyze how these dynamics affect production volatility and network resilience. Integrating Leontief input-output frameworks with controllability theory. By quantifying fluctuations as expected values under probabilistic node dynamics, we identify trade-offs between adaptability and stability. Methodologically, we unify Kalman rank criteria and minimum input theory, offering policymakers insights to balance innovation-driven entry with safeguards against destabilizing exits.
\end{abstract}

{\bf keywords: }networks, dynamic input-output, controllability, aggregate fluctuations
% \newpage
\section{Introduction}
Production networks, which consist of firms and industries interconnected through input-output relationships \cite{leontief1966input}, play a crucial role in determining how microeconomic shocks are transmitted throughout an economy. These networks act as conduits, transforming localized disturbances into widespread economic impacts, particularly in sectors with dense and asymmetric input-output linkages. An expanding body of research has focused on understanding how disruptions at the level of individual firms or industries propagate into broader economic fluctuations via these networks. Studies have shown that micro-level fluctuations can serve as a source of macroeconomic instability \cite{Gabaix2009}, with the structure of the network significantly influencing the magnitude of such fluctuations \cite{acemoglu2012}.
Recent research has increasingly utilized the Leontief input-output model to examine the behavior of production structures. In his seminal work on the American economy, Leontief emphasized the pervasive interconnectedness of industries, highlighting that even the most distant sectors of the economy are interlinked through input-output relationships. This model has proven useful in understanding the propagation of shocks across the economy. However, much of the existing literature assumes that these networks are static, with fixed nodes and connections, thus neglecting the dynamic nature of real-world markets where firms continually enter and exit \cite{Leontief1951}. Research on input-output networks explores their structure and evolution, highlighting universal features like fat-tailed weight distributions and clustering. \cite{LeonidovSerebryannikova2019} analyzed networks in Russia and the USA, revealing dynamic changes in metrics like centrality and edge weight slopes. Distributed input-output models address dynamic analysis challenges by accommodating lags, lead times, and singular matrices \cite{ThijsDynamicIO1986}. These models highlight how temporal investment distributions impact economic outcomes, as shown in studies of the Polish economy. \cite{TsyvinskiLiu2024} extend this approach by integrating input adjustment costs.

This article seeks to address this gap by proposing a novel approach to analyze macroeconomic fluctuations through the lens of probability, specifically focusing on the deletion and addition of nodes within input-output networks. While prior research has primarily examined the steady-state behavior of these networks, our approach emphasizes their dynamic evolution, where the entry and exit of firms are modeled as probabilistic events. We measure fluctuations in the economy by calculating the probability of node deletion and the addition of new nodes, considering these probabilities across the network. By incorporating these dynamic aspects, we offer a more nuanced and comprehensive framework for understanding how micro-level changes in the network structure—through the ongoing evolution of firms—affect the transmission of shocks and contribute to broader macroeconomic instability.We also introduce controllability as a criterion to measure the network's responsiveness to changes in a macroeconomic context. This approach evaluates how changes in node or edge dynamics affect the network's overall structure and its capacity to adapt.

A conceptual base model is first established to capture the dynamic interconnections between suppliers and customers within an economy, allowing for the measurement of spillover effects from individual firms to others, even through indirect linkages. These shocks are not external; rather, they are internal to the system, arising from the evolution of the input-output network itself. General equilibrium theory is employed to trace how localized microeconomic disturbances, driven by the ongoing evolution of firms within the network, propagate across the broader economy. By modeling the interactions between firms within an input-output graph, these tools provide a mechanism for understanding how aggregate economic fluctuations can be traced back to these internal, micro-level shocks. These tools offer valuable insights into the transmission of economic disturbances, illustrating how even small, firm-specific disruptions can cascade through the economy, ultimately affecting macroeconomic stability and contributing to broader fluctuations.

Second, after deriving the mathematical form of the influence matrix, the next step is to design a mechanism that explains the sequence of evolutionary steps within the network. This mechanism models the dynamics of network changes at each step, capturing how firms enter, exit, and adjust their positions within the input-output structure. Each stage of the network's evolution is crucial, as it shapes both the overall structure and the propagation of shocks across the economy. To further explore the long-term behavior of the network, this process is extended through a differential model that describes the probability distribution of node degrees in the stationary state. The model of the stationary state follows a scale-free structure, where the degree distribution exhibits power-law characteristics, reflecting the unequal distribution of interconnections typically observed in real-world networks. By combining these approaches, the model provides a comprehensive framework for analyzing both the short-term dynamics of network evolution and the long-term equilibrium behavior of the system, including how the distribution of firm interconnections stabilizes over time according to scale-free properties.

Third, The concept of controllability is central to understanding the dynamic behavior of the network, particularly in terms of guiding its evolution or mitigating disruptions. Two distinct methods are used to assess the controllability of the network \textit{Kalman Rank Criterion} the first method applies the Kalman rank criterion, which is particularly suited for probabilistic models. This method evaluates controllability by checking the rank of the controllability matrix derived from the network's dynamics, ensuring that all states of the network can be influenced by external inputs. \textit{Minimum Input Theory} the second method is based on minimum input theory, which focuses on determining the minimum number of control nodes required to steer the network. This approach provides insights into the long-term evolution of the graph, using results derived from differential equations to model the stationary state of the network. By combining these methods, the analysis not only assesses the current state of the network's controllability but also offers a predictive understanding of its long-term behavior. The Kalman rank method is ideal for short-term dynamics, while the minimum input theory aligns with the differential models governing the graph's evolution, providing a comprehensive view of the network's controllability across time scales.

Forth, we focus on obtaining the statistical behavior of fluctuations by employing numerical methods to analyze the random behaviors of certain parameters within the model, for which exact solutions are unavailable. Since some model parameters are inherently uncertain or not directly observable, we rely on stochastic simulations to approximate their distribution and assess their impact on network evolution. By generating multiple realizations of the system under different random configurations, we can derive statistical measures that capture the variability of fluctuations and the probabilistic nature of network changes. This approach enables us to explore the range of possible outcomes and gain insights into the behavior of the system under uncertainty, providing a more robust understanding of how fluctuations in the input-output network propagate through the economy.

In the following sections, this paper is organized as follows. Section \ref{sec:BaseModel} defines the model of production, outlining the dynamic interconnections between suppliers and customers within the economy and the mathematical formulation of the influence matrix. This section also establishes the framework for understanding how localized, microeconomic disturbances propagate through the input-output network. Section \ref{sec:EvolvingProduction} presents the mechanism that describes the evolution of the network, detailing how firms enter, exit, and adjust their positions within the network over time. It also explores the differential model used to analyze the probability distribution of node degrees in the stationary state, emphasizing the scale-free nature of the network structure. Section \ref{sec:MacroeconomicFluctuations} it examines the mechanisms through which productivity shocks within individual industries propagate across an economy, influencing aggregate output. Section \ref{sec:Controability} explains the controllability methods used to examine basic graphs and their evolved forms. Finally, Section \ref{sec:NumericalAnalyzing} focuses on the statistical behavior of fluctuations, employing numerical methods to capture the random behaviors of certain parameters within the model. Since exact solutions are not available for some of these parameters, stochastic simulations are used to estimate the probabilistic impact of network changes on macroeconomic fluctuations.

\section{Base model}
\label{sec:BaseModel}
In this section, we introduce a model of production networks that incorporates the dynamic processes of node addition and deletion. Building on the foundational frameworks of \cite{acemoglu2012} and \cite{TahbazCarvalho2019}, our model extends traditional static input-output linkages, which focus on the propagation of shocks within fixed network structures. To reflect real-world dynamics more accurately, we propose a dynamic extension of the production function that accounts for firm entry (node addition) and exit (node deletion), enabling structural changes in the network over time.

\subsection{Production network} \label{sec:ProductionNetwork}

Consider an economy consisting of \( n \) sectors, indexed by \( i \in \{1, 2, \dots, n\} \), where each sector produces a distinct product. The output of sector \( i \), denoted \( y_i \), is determined by a Cobb-Douglas production function that combines labor and intermediate inputs from other sectors. Specifically, the production function for sector \( i \) is given by:

\begin{equation}
    y_i = \beta_i \psi_i L_i^{\alpha_i} \prod_{j} x_{ij}^{a_{ij}},
    \label{eq:production_function}
\end{equation}

where \( \beta_i \) is a sector-specific productivity parameter, \( \psi_i \) represents an internal shock to sector \( i \) that affects its productivity, \( L_i \) is the labor input in sector \( i \), and \( x_{ij} \) is the intermediate input from sector \( j \) used in the production of \( y_i \). The parameters \( \alpha_i \) and \( a_{ij} \) represent the output elasticities with respect to labor and the intermediate input from sector \( j \), respectively. The inclusion of \( \psi_i \) allows for the impact of internal shocks on the production capacity of each sector, which in turn affects both the output of the sector and the intersectoral relationships as inputs are transmitted throughout the network.
The variable \( \psi_i \), which captures the effects of node addition and deletion, is given by:

\begin{equation}
    \psi_i = \frac{\prod_{p} \left( x_{ip}^{b_{ip}} \right)}{\prod_{k} \left( x_{ik}^{c_{ik}} \right)}
    \label{eq:psi_variable}
\end{equation}

Here, \( x_{ip}^{b_{ip}} \) represents the quantity of intermediate inputs from newly added firms, while \( x_{ik}^{c_{ik}} \) captures the quantity of inputs from firms that have exited the network. The elasticities \( b_{ip} \), \( c_{ik} \), and \( a_{ij} \) denote the shares of inputs from new, exiting, and existing firms, respectively. The term \( \psi_i \) acts as an internal shock variable, reflecting structural changes in the network due to the entry and exit of firms. When \( \psi_i > 1 \), this indicates a net increase in inputs from newly added firms, whereas when \( \psi_i < 1 \), it implies reduced input availability due to firm exits. Thus, \( \psi_i \) measures how internal changes within the network affect industry \( i \)'s production capacity.
The variables \( x_{ip}^{b_{ip}} \), \( x_{ik}^{c_{ik}} \), and \( x_{ij}^{a_{ij}} \) are defined over a continuous domain, where the quantities of them vary smoothly with firm entry/exit. Specifically, we define \( x_{ip}^{b_{ip}}, x_{ik}^{c_{ik}}, x_{ij}^{a_{ij}} : \mathbb{R}^+ \to \mathbb{R}^+ \), where \( \mathbb{R}^+ \) denotes the set of positive real numbers. This continuous formulation allows the model to track the flow of inputs over time, reflecting gradual changes in the composition of firms and their input contributions.

In the base model, the relationship between \( \psi_i \) and the production output \( y_i \) is treated as a direct interaction, where \( \psi_i \) influences the production capacity by modifying the availability of inputs from different firm cohorts. However, in the extended model, this relationship becomes probabilistic. Specifically, \( \psi_i \) is now used to calculate the likelihood of different scenarios regarding the entry and exit of firms, which in turn affects the expected value of inputs in the production function. In this context, \( \psi_i \) no longer just shifts production capacity but also serves as a parameter that influences the probability distribution of input availability. The expected value of \( y_i \) is thus computed by incorporating these probabilities, allowing for a more nuanced modeling of how firm dynamics—such as entry, exit, and structural shifts—affect overall production outcomes.

To maintain the standard Cobb-Douglas form, we impose a constraint requiring the sum of elasticities across all inputs and labor to equal one:
\begin{equation}
    \alpha_i + \sum_{p} b_{ip} - \sum_{k} c_{ik} + \sum_{j} a_{ij} = 1
    \label{eq:elasticity_constraint}
\end{equation}

This constraint ensures constant returns to scale, with the negative sign on \( \sum_{k} c_{ik} \) reflecting the reduction in output due to the absence of inputs from deleted firms. Specifically, \( \alpha_i \), \( b_{ip} \), \( c_{ik} \), and \( a_{ij} \) represent the shares of output attributable to labor, inputs from newly added firms, inputs from firms that have exited, and inputs from existing firms, respectively. All these variables are bounded within the interval \( [0, 1] \). Additionally, this constraint helps in the simplification of the resulting equations, as it allows for a reduction in the number of variables needed to describe the system, making the model more tractable and easier to solve. Alongside the firms described earlier, the economy also includes a representative household that supplies one unit of labor inelastically and exhibits logarithmic preferences over the \( n \) goods, expressed as:

\begin{equation}
    u(c_1, c_2, \ldots, c_n) = \sum_{i=1}^n \gamma_i \log\left(\frac{c_i}{\gamma_i}\right)
    \label{eq:utility_function}
\end{equation}
where \( c_i \) is the amount of good \( i \) consumed. The constants \( \gamma_i \geq 0 \) measure various goods' shares in the household's utility function, normalized such that \( \sum_{i=1}^n \gamma_i = 1 \).

The environment is fully defined by the system of equations in \eqref{eq:production_function}, \eqref{eq:psi_variable}, \eqref{eq:elasticity_constraint}, and \eqref{eq:utility_function}. In the competitive equilibrium of this economy, the following conditions hold: (i) the representative household optimizes its utility by choosing consumption; (ii) each representative firm in the sectors maximizes its profit, given the prices and wage rate; and (iii) all markets clear, ensuring that supply equals demand in every market.

The addition and deletion of nodes affect the overall production output in several ways. When a new node is added (e.g., a new firm or sector enters the economy), it increases the diversity of available inputs, represented by \( x_{ip} \). This can increase the output \( y_i \) by providing more intermediate inputs or reducing reliance on a small number of suppliers. On the other hand, when a node is deleted (e.g., a firm exits the market), represented by \( x_{ik} \), the production output is negatively affected because the network loses a source of input, potentially leading to bottlenecks or shortages.

Thus, the inclusion of both node additions and deletions in the production function allows us to capture the dynamic nature of real-world economies, where firms frequently enter and exit the market. These dynamics are particularly important for understanding how shocks propagate through the network. For instance, the deletion of a major supplier (a critical node) can have disproportionate effects on industries that depend heavily on that supplier, leading to cascading failures across the network. \cite{Baqaee2018cascading} explores how firm entry and exit in production networks amplify idiosyncratic shocks by altering input-output relationships. His model shows that firm exits in critical industries can significantly disrupt equilibrium output, highlighting how structural changes in the network affect systemic risk. Similarly, \cite{Elliott2022Fragility} focuses on network fragility, where the interconnectedness of agents—modeled through network dependencies and phase transitions—makes the system highly sensitive to localized shocks. Their approach quantifies the potential for widespread disruptions based on these structural features. However, in this model, we introduce a novel method that measures the risk of such disruptions more directly, based on the structure of the network itself, allowing for a more nuanced understanding of how changes in network topology influence the propagation of shocks and, ultimately, aggregate shocks.

In equilibrium, firms in each industry \( i \) maximize profits by choosing the optimal amounts of labor \( L_i \) and intermediate inputs \( x_{ij}, x_{ip}, x_{ik} \), given the prices of inputs and outputs. The profit function for industry \( i \) is given by:

\begin{equation}
\pi_i = p_i y_i - w L_i - \sum_j p_j x_{ij} - \sum_p p_p x_{ip} + \sum_k p_k x_{ik}
\label{eq:profit_function}
\end{equation}

where \( p_i \) is the price of the output in industry \( i \), \( w \) is the wage rate for labor \( L_i \), \( p_j \) is the price of intermediate input \( x_{ij} \) from existing firms, \( p_p \) is the price of intermediate input \( x_{ip} \) from newly added firms (node additions), and \( p_k \) is the price of intermediate input \( x_{ik} \) from firms that have exited the network (node deletions). The term \( + \sum_k p_k x_{ik} \) reflects the benefit from inheriting inputs from exiting firms.

To find the optimal values of labor and inputs, we substitute the production function given in equation \eqref{eq:production_function} into the profit function specified in equation \eqref{eq:profit_function}. We then derive the first-order conditions by taking the partial derivatives of the profit function with respect to each variable—labor and inputs. By setting each of these derivatives equal to zero, we determine the points where further changes in any variable would no longer increase profits, thereby identifying the optimal values for firms. Based on this process, the optimal input quantities are given by \( x_{ik} = p_i c_{ik} y_i / p_k \), \( x_{ip} = p_i b_{ip} y_i / p_p \), and \( x_{ij} = p_i a_{ij} y_i / p_j \), while the optimal amount of labor is \( L_i = p_i \alpha_i y_i / w \).

\begin{equation}
\log{(p_i/w)} = \sum_j{a_{ij}\log{(p_j/w)}}+\sum_j{b_{ip}\log{(p_p/w)}}-\sum_j{c_{ik}\log{(p_k/w)}}
\label{eq:log_price_wage}
\end{equation}

\subsection{Step Matrices}

The matrices \( \mathbf{A} \), \( \mathbf{B} \), and \( \mathbf{C} \) represent the internal structure of the economic system. Specifically, \( \mathbf{A} \) captures the primary input-output relationships between firms and industries, where the coefficients \( a_{ij} \) describe the interactions between industry \( i \) and industry \( j \). Matrix \( \mathbf{B} \) is associated with the introduction of new nodes into the system, with coefficients \( b_{ip} \) indicating dependencies between industries and intermediate products that arise from the addition of new nodes. In contrast, matrix \( \mathbf{C} \) is related to the deletion of nodes from the network, where coefficients \( c_{ik} \) represent the influence of input categories that are lost due to node deletions.
Rewriting the system of equations in matrix form, we have:

\begin{equation}
\hat{p} = \mathbf{A} \hat{p} + \mathbf{B} \hat{p} - \mathbf{C} \hat{p},
\label{eq:relative_price}
\end{equation}
where \( \mathbf{A} \), \( \mathbf{B} \), and \( \mathbf{C} \) together characterize the internal structure of the economy. In this equation, \( \hat{p} \) is the vector of log relative prices, defined as:

\begin{equation}
\hat{p} = \left( \log\left(\frac{p_1}{w}\right), \log\left(\frac{p_2}{w}\right), \dots, \log\left(\frac{p_n}{w}\right) \right)',
\label{eq:relative_price_vector}
\end{equation}
where \( p_i \) represents the price of good \( i \), and \( w \) is a reference price level (e.g., the wage rate or a common price index). This matrix formulation allows us to solve for all relative prices based on the underlying structure of the economy.

To streamline the solution process, we define a new matrix \( \mathbf{E} = \mathbf{B} - \mathbf{C} \), which captures the net effect of matrices \( \mathbf{B} \) and \( \mathbf{C} \) on the system, reflecting the combined influence of node additions and deletions. The elements of \( \mathbf{E} \) are treated as random variables, reflecting the stochastic nature of the network's evolution, such as the probabilistic patterns of node additions and deletions. Thus, the distribution of each element \( e_{ij} \) in \( \mathbf{E} \) is influenced by the probabilistic patterns of these network modifications. The solution for \( \hat{p} \) can now be expressed as:
\begin{equation}
\hat{p} = - \left(\mathbf{I} - (\mathbf{A} + \mathbf{E})\right)^{-1}.
\label{eq:dynamic_inverse_leontief}
\end{equation}

with statistical analysis of each changes along the steps of times this model try to get rate of addition and deletion nodes as initial condition then normalize them with one node addition. Consider $m$ is addition rate and $n$ is deletion rate is defined $m, n \in \mathbb{Q}$, such that $m > n$ normal forms are 1 addition rate and \(n/m\) deletion rate. As an exmaple when our dynamic network addition node is 2 and 1 deletion node rate by normilization respect to addition node we get 1 node additon and 0.5 node deletion in each step\footnote{Because of deletion and addition are in the same time we consider deletion is just for exist node and new column and row which is added should be zero.}.

In the context of scale-free networks, the mechanism of preferential attachment governs the connection probabilities between nodes \cite{BarabasiScaleFree1999}. Specifically, new nodes are more likely to attach to existing nodes that already have a higher degree (i.e., more connections). This process results in the formation of hubs, or nodes with many connections, while other nodes have fewer connections \cite{chung2006complex}, \cite{clauset2009power}. 

The probability \( P_i \) of a new node attaching to an existing node \( i \) is proportional to the degree \( k_i \) of that node:
\begin{equation}
    p_i = \frac{k_i}{\sum_j{k_{ij}}}
    \label{eq:p_addition}
\end{equation}
where \( k_i \) represents the degree of node \( i \). This preference for higher-degree nodes drives the emergence of a power-law degree distribution in the network, characteristic of scale-free networks. Conversely, the probability of a node being deleted or disconnected from the network is inversely related to its degree.
\begin{equation}
    p_{d_i} = \frac{1/k_i}{\sum_j{1/k_{ij}}}
    \label{eq:p_deletion}
\end{equation}
This deletion process typically affects lower-degree nodes more severely, as they are less likely to attract new connections. As a result, low-degree nodes are more vulnerable to removal, leading to a network where the deletion or disconnection of nodes follows a pattern that depends on the degree distribution of the network.

This formulation ensures that the matrix is appropriately dimensioned for the analysis of dynamic evolution and control systems. The concept of the \textit{step matrix} derives from the dynamic extension of the Leontief input-output framework, as proposed by \cite{ThijsDynamicIO1986} in his work on dynamic input-output analysis. This extension serves as a foundation for incorporating the \textit{step matrix} into control theory applications \cite{Vogt1975dynamicLeontief}. Additionally, it plays a crucial role in growth analysis models, such as those explored by \cite{Kurz2000DynamicLeontiefModelGrowth}, which apply dynamic Leontief models to examine economic growth.

The \textit{step matrixes} is defined in the shape of below, matrix $\mathbf{A}$ get zero column and row and prepared for to be correct dimension of summation in countinue.
\[
\mathbf{A} = \begin{pmatrix}
    a_{11} & a_{12} & \cdots & a_{1,n-1} & 0      \\
    a_{21} & a_{22} & \cdots & a_{2,n-1} & 0      \\
    \vdots & \vdots & \ddots & \vdots & \vdots \\
    a_{m-1,1} & a_{m-1,2} & \cdots & a_{m-1,n-1} & 0      \\
    0      & 0      & \cdots & 0      & 0
\end{pmatrix}
\]
matrix $\mathbf{B}$ shows node addition in each step then by this definition we have Expected value of probability of addition node to each exist nodes.
\[
\mathbf{B} = \begin{pmatrix}
    b_{11}              & b_{12}            & \cdots             & b_{1,n-1}            & \mathbb{E}[b_{1n}]\\
    b_{21}              & b_{22}            & \cdots             & b_{2,n-1}            & \mathbb{E}[b_{2n}]\\
    \vdots              & \vdots            & \ddots             & \vdots            & \vdots            \\
    b_{m-1,1}              & b_{m-1,2}            & \cdots             & b_{m-1,n-1}            & \mathbb{E}[b_{4n}]\\
    \mathbb{E}[b_{m1}]  &\mathbb{E}[b_{m2}] & \cdots             &\mathbb{E}[b_{m4}] & \mathbb{E}[b_{mn}]
\end{pmatrix}
\]
It could have been defined all elements of $\mathbf{B}$ zero ($b_{ij}=0$) except $\mathbb{E}[b_{im}]$ and $\mathbb{E}[b_{mj}]$,  but because of generality of model consideration of (\ref{eq:elasticity_constraint}) must also be taken into account. Matrix $\mathbf{C}$ is the effect of node deletion probability according to \( 1 - p_{i} \) and considered $ p_i \propto k_{i} $. Also new nodes is not in the deletion probabilities and set \( p_{i} = 1 \) then \(1-p_{i}\) is equal zero. Elements of matrix is defined $ c_{ij} = p_i.a_{ij} $:
\[
\mathbf{C} = \begin{pmatrix}
    c_{11} & c_{12} & \cdots & c_{1,n-1} & 0\\
    c_{21} & c_{22} & \cdots & c_{2,n-1} & 0\\
    \vdots & \vdots & \ddots & \vdots & 0\\
    c_{m-1,1} & c_{m-1,2} & \cdots & c_{m-1,n} & 0\\
    0      & 0      & 0      & 0      & 0
\end{pmatrix}
\]
Using this formulation, the Leontief inverse \( \mathbf{L} \) is updated as \cite{stewart1998matrix}:
\begin{equation}
\mathbf{L} = (\mathbf{I} - \mathbf{A} - \mathbf{E})^{-1} = \sum_{k=0}^{\infty} (\mathbf{A} + \mathbf{E})^k.
\label{eq:leontief_with_p}
\end{equation}
For each element \( l_{ij} \) in \( \mathbf{L} \), we obtain:
\begin{equation}
l_{ij} = a_{ij} + \mathbb{E}[e_{ij}] + \sum_{r=1}^{n} (a_{ir} + \mathbb{E}[e_{ir}]) (a_{rj} + \mathbb{E}[e_{rj}]) + \ldots,
\label{eq:lij_with_p}
\end{equation}
where \( \mathbb{E}[e_{ij}] \) is replaced by elements of matrixes which is defined, as definition in step matrix.
The first term in Equation \eqref{eq:lij_with_p} represents the direct effect of \( i \) on \( j \), while subsequent terms incorporate indirect effects mediated through intermediate nodes. These indirect effects include both deterministic contributions from \( \mathbf{A} \) and stochastic contributions via the probability-weighted \( \mathbb{E}[e_{ij}] \), computed based on the probability of addition and deletion.

\section{Evolving production networks}
\label{sec:EvolvingProduction}
In the context of firm networks, step matrices offer a powerful tool for modeling changes across successive evolutionary stages, transitioning from time \( t \) to \( t+1 \). As \( t \to \infty \), this dynamic evolution, studied through differential methods, allows for the identification of stationary equilibria within the system. A stationary equilibrium represents the long-term transition probabilities, which are inherently determined by the degree distribution of the network's nodes. Specifically, the degree of a node—representing the number of buyer-supplier connections a firm maintains—is directly influenced by the network's overall structure, including the number of edges and nodes. Understanding the relationship between network growth and degree distribution is essential for analyzing how stationary degree distributions emerge as the system reaches a state of stability over time.

Recent research on the evolution of complex networks has provided valuable insights into the dynamics of networks with both node addition and deletion processes. In their work, \cite{Moore2006ExactSF} developed a model in which networks evolve through preferential attachment, where new nodes connect to existing ones with a probability proportional to their degree. This mechanism results in power-law degree distributions, with the exponent varying as a function of the node addition rate. Such models are particularly useful in understanding network evolution where both growth and decay processes are present, reflecting more realistic features of real-world systems. Similarly, \cite{BenNaim2007AdditiondeletionN} extended this framework to analyze evolving networks with node addition and deletion. Their model provides analytical solutions for the in-component size distribution and other structural properties of the network, shedding light on the long-term behavior of networks under these dual processes.

\cite{Atalay2011} applied similar principles to the buyer-supplier network of the US economy, demonstrating that purely scale-free models fail to capture the key characteristics of this network. By incorporating more realistic features of firm-level buyer-supplier relationships, they constructed an alternative model that better aligns with empirical data. This model, which builds on the preferential attachment framework of \cite{Moore2006ExactSF}, offers a more accurate representation of the network's structure and enhances the understanding of important economic phenomena. 

Given the alignment between the models presented by \cite{Moore2006ExactSF}, \cite{BenNaim2007AdditiondeletionN}, and \cite{Atalay2011}, particularly in their use of preferential attachment and the incorporation of both addition and deletion processes, we adopt a similar framework for analyzing the stationary equilibrium in firm networks. This model, rooted in the dynamics of network growth and decay, allows for the exploration of the long-term equilibrium degree distributions as the network evolves over time and stabilizes. By leveraging this approach, we can better understand how the firm network reaches a stationary state and how its structure and degree distribution evolve as time progresses to infinity.

Consider a network of firms, where each node corresponds to a firm and each edge represents a buyer-supplier relationship. These relationships evolve continuously through processes such as firm entry and the rewiring of connections, which collectively drive the dynamic reshaping of the network's structure. Let \( p(k, t) \) denote the density of firms with degree \( k \) at time \( t \), where \( k \) represents the number of suppliers associated with a firm. The temporal evolution of \( p(k, t) \), reflecting the changing distribution of firm degrees over time, is governed by the following partial differential equation:

\begin{equation}
    \frac{\partial p(k, t)}{\partial t} + \frac{\partial}{\partial k} \left[ \phi(k, t)p(k, t) \right] = M(k, t)(q + g) - (q + g)p(k, t),
    \label{eq:diff_equation_of_evolve}
\end{equation}

Here, \( \phi(k, t) \) is the degree growth rate, \( M(k, t) \) is the rate of new firms entering the market, \( q \) is the exit rate of firms, and \( g \) denotes the network's net growth rate. This mathematical model has been successfully solved and validated in the study by \cite{Atalay2011}.

In the steady-state regime (\( t \to \infty \)), the stationary degree distribution \( p(k) \) satisfies the following equation:

\begin{equation}
\frac{\partial}{\partial k} \left[ p(k) \left( qr(\nu - k) + \frac{\delta(k + r(\nu - k))(q + g)}{1 - q} \right) \right] = (q + g) \left( \frac{e^{-k / [\nu(1 - \delta)]}}{\nu(1 - \delta)} - p(k) \right),
\label{eq:diff_equation_of_evolve_2}
\end{equation}
In this equation, \( \nu \) is the average degree, \( \delta \) represents the fraction of edges connecting new nodes to existing firms, and \( r \) is the proportion of rewired edges distributed uniformly across the network. This equation captures the dynamics governing the evolution of the degree distribution, accounting for firm exits, edge rewiring, and preferential attachment mechanisms.

The analytical solution for the stationary degree distribution is given by:
\begin{equation}
p(k) = \lambda (k + R)^{-1 - S} \left[ \Gamma\left(1 + S, \frac{R}{\nu(1 - \delta)} \right) - \Gamma\left(1 + S, \frac{R + k}{\nu(1 - \delta)} \right) \right],
\label{eq:evolveProb}
\end{equation}
where \( \Gamma(a, x) \) is the upper incomplete gamma function. The parameters \( R \) and \( S \) are defined as:
\begin{equation}
R = \frac{\nu \delta (q + g) r + q r (1 - q)}{\delta (q + g)(1 - r) - q r (1 - q)}, \quad S = \frac{(q + g)(1 - q)}{\delta(1 - r)(q + g) - q r (1 - q)}.
\label{eq:probparamRS}
\end{equation}
Here, \( \lambda \) is a normalization constant ensuring that the degree distribution sums to one, explicitly defined as:
\begin{equation}
\lambda = \exp{\left (\frac{R}{\nu(1 - \delta)} \right )} S [\nu(1 - \delta)]^S.
\label{eq:probparamlambda}
\end{equation}
The parameters \( R \) and \( S \) encapsulate the interplay between network growth dynamics, preferential attachment, and edge rewiring mechanisms.

The observed behavior of \( p(k) \) suggests a remarkably consistent pattern, indicative of an underlying structural regularity. The distribution's form aligns with characteristics commonly associated with power-law behavior, a phenomenon frequently encountered in economic and social systems where underlying network dynamics or proportional growth mechanisms drive outcomes. Such forms not only capture tail-heavy distributions but also provide insights into scaling properties and the concentration of economic activities or resources.

It seems, therefore, that \( p(k) \) adheres strongly to the attributes of a power function. In the subsequent numerical analysis section, we will approximate this relationship by modeling \( p(k) \) as \( k^{-\kappa} \), allowing us to explore its implications quantitatively and to derive parameter estimates that further illuminate the economic significance of this distribution.

The probability distribution outlined in Equation~\ref{eq:evolveProb_new} encapsulates the principle of preferential attachment, where the likelihood of establishing new connections is directly linked to a firm's current degree. This mechanism illustrates the ``rich-get-richer'' effect, demonstrating that firms with a greater number of existing connections are more prone to acquiring additional links over time.

By concentrating exclusively on preferential attachment and omitting uniform edge rewiring, our model highlights the critical role of degree-dependent link formation in determining the structural characteristics of firm networks. This deliberate exclusion facilitates a more focused examination of how preferential attachment alone drives network dynamics, offering deeper insights into the concentration of relationships among leading firms. As a result, this framework provides a refined depiction of buyer-supplier interactions, enhancing the understanding of the fundamental processes that govern network evolution within the U.S. economy.

\begin{equation}
    P_a(k) = \frac{k}{(1 - q)N(t)\nu(t)}
\label{eq:evolveProb_new}
\end{equation}
Here, \( P_a(k) \) represents the probability that a new link attaches to a firm with degree \( k \), adjusted to accommodate our model, which includes one unit of addition and \( {n}/{m} \) deletions. Consequently, the preferential attachment probability is defined as
\begin{equation}
    P_a(k) = \frac{k}{\left(1 - \frac{n}{m}\right)t}
    \label{eq:evolveProb_new2}
\end{equation}

\section{Network-Originated Macroeconomic Fluctuations}
\label{sec:MacroeconomicFluctuations}
This section explores the link between macroeconomic volatility and internal shocks, referencing key literature. \cite{Hulten1978} refines growth accounting by emphasizing intermediate inputs, showing that traditional methods often underestimate the impact of technological progress by ignoring the relationship between productivity and intermediate goods accumulation. \cite{acemoglu2012} challenge the view that firm-level shocks average out, proposing the granular hypothesis, which argues that large firms in economies with fat-tailed distributions can significantly affect overall economic performance. While \cite{lucas1977} and others have downplayed the role of microeconomic shocks, citing the ``diversification argument,'' which assumes that independent shocks in many industries reduce aggregate fluctuations, this view overlooks the interconnectedness between sectors. Shocks can propagate through input-output linkages, leading to correlated effects across industries and amplifying macroeconomic volatility.

Two fundamental theorems are presented here that elucidate how productivity shocks within individual industries are propagated through an economy and influence aggregate output, by using the baseline model in Section \ref{sec:BaseModel}. The analysis is based on the Leontief inverse matrix and Domar weights within a Cobb-Douglas framework to capture both direct and indirect effects of such shocks.By applying the methodology of \cite{TahbazCarvalho2019}, we obtain the following:
\newline\newline
\textbf{Theorem 1:} \label{th:1}
Indicates that the output of an industry is influenced by productivity shocks that propagate "downstream" from an industry to its customers, customers' customers, and so on. This propagation is captured by the economy's Leontief inverse, which measures the importance of an industry as a direct and indirect input supplier to another industry.

\begin{equation}
\log(y_i) = \sum_{j=1}^n l_{ij}\label{log_y_i}
\end{equation}
where \( l_{ij} \) is given by equation \eqref{eq:lij_with_p} which incorporates internal shocks in its calculation.
\newline\newline
\textbf{Theorem 2:} \label{th:2}
Shows that an industry's Domar weight, which represents its sales as a fraction of GDP, is a sufficient statistic for how its shocks impact aggregate output.

\begin{equation}
\log(\text{GDP}) = \sum_{j=1}^n \lambda_{ij}\label{log_gdp}
\end{equation}
where 
\begin{equation}
    \lambda_i = \frac{p_iy_i}{\text{GDP}} = \sum_{j=1}^n \gamma_i l_{ij} \label{lambda}
    \end{equation}
\( l_{ij} \) is given by equation \eqref{eq:lij_with_p} which incorporates internal shocks in its calculation. Theorem 2 shows that with Cobb-Douglas preferences and technologies, Domar weights depend on preference shares and the Leontief inverse. This means that shocks to more important industries have a larger impact on aggregate output, as their increased Domar weight amplifies the downstream effects.

The model by \cite{acemoglu2012} provides a foundational approach to understanding aggregate fluctuations arising from sectoral shocks within production networks. Specifically, this analysis derives an aggregate volatility formula, showing that idiosyncratic shocks in a network of interconnected industries, when propagated through production linkages, can culminate in significant fluctuations at the aggregate level. This result challenges the traditional view that such microeconomic disturbances would average out at higher levels of disaggregation. It demonstrate that the asymptotic behavior of aggregate volatility is strongly influenced by the structural properties of the production network, notably the distribution of industry centralities.

In particular, The relationship is characterized, who analyze how the decay rate of aggregate volatility varies depending on whether the network exhibits symmetry in its input-output linkages or contains sectors that disproportionately serve as central suppliers. It is shown in the model that, in symmetric networks, aggregate volatility decays at the rate of $1/\sqrt{n}$, consistent with the law of large numbers. However, in more asymmetric configurations, such as star networks, sector-specific shocks propagate extensively, leading to much slower decay rates of aggregate volatility.

In this study, we extend the framework of \cite{acemoglu2012} and \cite{TahbazCarvalho2019} by focusing on internal shocks within the network, represented by $\psi_i$, where each $\psi_i$ is a continuous random variable with standard deviation $\sigma = \text{std}(\psi_i)$. The aggregate output volatility, $\sigma_{\text{agg}}$, is defined as the standard deviation of the log of GDP:
\begin{equation}
    \sigma_{\text{agg}} = \text{std}(\log(\text{GDP})) = \sigma \, \|\lambda\|,
\end{equation}
where $\|\lambda\|$ represents the norm of the vector of Domar weights:
\begin{equation}
    \|\lambda\| = \sqrt{\sum_{i=1}^n \lambda_i^2}.
\end{equation}
Each Domar weight $\lambda_i$ is calculated as:
\begin{equation}
    \lambda_i = \frac{v_i}{n},
\end{equation}
with $v_i = \sum_{j=1}^n \ell_{ji}$ capturing the centrality of industry $i$ as an input supplier.

A significant addition to our analysis is a mechanism of node addition and deletion to dynamically model how shifts in network structure influence the distribution of centralities, $v_i$, and thereby impact aggregate volatility, $\sigma_{\text{agg}}$. This mechanism is particularly relevant for assessing the sensitivity of $\sigma_{\text{agg}}$ in networks where industry centralities exhibit Pareto-distributed heavy tails. In such networks, aggregate volatility scales with the rate $n^{1/\gamma - 1}$, indicating that highly asymmetric structures can experience amplified volatility well beyond the standard $1/\sqrt{n}$ rate.

The relationship between $\sigma_{\text{agg}}$ and the network's structure of input-output linkages is thus expressed as:
\begin{equation}
    \sigma_{\text{agg}} = \frac{\sigma}{\sqrt{n}} \sqrt{\alpha^{-2} + \text{var}(v_1, \ldots, v_n)},
    \label{eq:aggregate_volatility}
\end{equation}
where $n$ denotes the number of industries, $\alpha$ the labor share, and $\text{var}(v_1, \ldots, v_n)$ the variance in centralities.

The structural impact of node addition and deletion is further highlighted through the modified Leontief inverse:
\begin{equation}
    L = I + L(A + B - C),
\end{equation}
where $A$, $B$, and $C$ represent input-output matrices incorporating dependencies beyond simple production linkages.

The network's asymmetry plays a critical role in determining whether shocks cancel out or propagate. In symmetric networks, internal shocks $\psi_i$ tend to offset each other, minimizing $\sigma_{\text{agg}}$. In contrast, asymmetric networks, such as those with star-like configurations, can exhibit significant amplification in $\sigma_{\text{agg}}$ due to the high variance in centralities $v_i$. Our node addition and deletion mechanism allows for modeling shifts in network asymmetry, where adding highly connected nodes increases the skew in the $v_i$ distribution, potentially heightening $\sigma_{\text{agg}}$. Conversely, removing central nodes or adding evenly distributed ones mitigates aggregate volatility by promoting a more balanced centrality distribution.

\section{Controability}
\label{sec:Controability}
In a fluctuating state within production or supply chain systems, it is essential for governments and major corporations to regulate and manage macroeconomic parameters or introduce policies that mitigate the spread of volatility. Such measures are crucial to prevent systemic failures within these networks. Recent research has highlighted the network structure of the U.S. economy, employing mathematical proof methods to analyze these networks in depth, as discussed in Section \ref{sec:EvolvingProduction}.

The classical concept and conditions of controllability have traditionally been applied to undirected networks. While controlling undirected networks presents its own intellectual challenges, these methods have found limited application in complex systems. This limitation arises because, to the best of our knowledge, most real-world complex networks where control is impactful are directed in nature. This distinction significantly restricts the practical applicability of previous theoretical results.

From a computational perspective, earlier methods encounter significant challenges when applied to large networks. First, identifying the minimum set of driver nodes requires testing configurations that scale exponentially with the network size, \(O(2^N)\), rendering the process computationally infeasible for networks containing thousands or millions of nodes. Second, applying Kalman's rank condition, a widely used criterion for controllability, becomes increasingly impractical at such scales. Consequently, most prior studies have focused on evaluating controllability using a limited set of small graph models, typically containing only a few dozen nodes—far smaller than the large-scale networks observed in real-world complex systems.

Furthermore, many prior studies assume that interaction strengths or link weights can be measured with complete accuracy. Some even make the simplifying assumption that all links have identical weights. These assumptions transform the controllability problem into a spectral graph theoretic problem, such as analyzing the spectrum of the Laplacian matrix of the network. While this simplification has allowed progress, it is often unrealistic in real-world networks where measurement errors and uncertainties are unavoidable. For instance, in regulatory networks, which are typically directed and weighted, we currently lack tools to accurately estimate link weights. This inherent uncertainty further complicates the application of traditional controllability methods to real-world networks.

Despite these computational challenges, Kalman's rank condition continues to serve as a foundational framework for evaluating controllability, particularly in the context of well-defined, small-scale network models, where its application provides valuable insights into system dynamics.

To address the limitations associated with scaling and network complexity, this study employs the Minimum Input Theory in Structural Control. This framework provides robust analytical tools capable of analyzing directed networks of arbitrary size and accommodating networks with heterogeneous and arbitrary weight distributions \cite{Liu2011}.

\subsection{Kalman Rank Condition for Controllability} 

Consider a linear time-invariant (LTI) system defined as:

\[
\dot{x}(t) = A x(t) + \hat{E} u(t),
\]

where:
\begin{itemize}
    \item \(x(t) \in \mathbb{R}^n\) is the state vector,
    \item \(u(t) \in \mathbb{R}^m\) is the control input vector,
    \item \(A \in \mathbb{R}^{n \times n}\) is the state matrix,
    \item \(\hat{E} \in \mathbb{R}^{n \times m}\) is the input matrix.
\end{itemize}

The system is \textbf{controllable} if and only if the \textit{controllability matrix}:

\[
\mathcal{C} = \begin{bmatrix} \hat{E} & A\hat{E} & A^2\hat{E} & \cdots & A^{n-1}\hat{E} \end{bmatrix}
\]

has full rank, i.e.,

\[
\text{rank}(\mathcal{C}) = n,
\]

where:
\begin{itemize}
    \item \(\mathcal{C} \in \mathbb{R}^{n \times (mn)}\) is the controllability matrix,
    \item \(n\) is the dimension of the state vector.
\end{itemize}

The elements of \(\hat{E}\) can be functions of the degrees of the nodes within the network. Here, the \textbf{degree} of a node refers to the number of connections (edges) it has with other nodes in the network. Let us consider a network represented by a graph \(G = (V, E)\), where \(V\) is the set of nodes and \(E\) is the set of edges. The degree of node \(i\), denoted by \(k_i\), plays a significant role in determining the influence of that node on the overall system dynamics through the addition and deletion of expected values. We define \(\hat{E}\) as a \textbf{diagonal matrix}. This approach simplifies the influence of each control input by ensuring that each control input affects only a specific node directly, without immediate cross-influence on other nodes. Formally, the input matrix \(\hat{E}\) can be expressed as:
\[
\hat{E} = \text{diag}\left(f(k_1), f(k_2), \ldots, f(k_n)\right),
\]
where:
\begin{itemize}
    \item \(\text{diag}(\cdot)\) denotes a diagonal matrix with the provided elements on its main diagonal.
    \item \(f(k_i)\) is a function that maps the degree \(k_i\) of node \(i\) to the corresponding input influence. This function determines how strongly each node is influenced by its respective control input based on its degree.
\end{itemize}
Defining \(\hat{E}\) as a diagonal matrix offers several advantages in the context of controllability. Firstly, it provides simplicity of control inputs, as each control input \(u_j(t)\) affects only one node \(i\), thereby simplifying the design and analysis of control strategies. Secondly, a diagonal \(\hat{E}\) promotes enhanced sparsity, which is beneficial for large-scale networks by reducing computational complexity and improving scalability. Lastly, it establishes a clear association between inputs and nodes by directly correlating control inputs with specific nodes, making it easier to understand and manage the influence each input has on the network.

\subsection{Minimum Input Theory in Structural Control}

The minimum matching model for network controllability is highly effective because it provides a clear, structured framework for identifying the minimum number of external inputs (driver nodes) required to control a directed network. By focusing on the concept of maximum matching, the model ensures that the network is efficiently managed, balancing internal control via matched nodes with external inputs for unmatched nodes. This approach is particularly powerful for directed networks, where influence flows in specific directions, making traditional undirected network methods insufficient. Its reliance on structural properties rather than exact link weights makes it robust in real-world applications, where uncertainties and measurement errors are common. This model is especially suitable for measuring the controllability of random graphs based on probability, as it leverages statistical physics methods to analyze networks with varying degree distributions. For instance, using tools like the cavity method, the model can determine the average number of unmatched nodes in ensembles of random graphs with prescribed probabilities for edge formation. This makes it possible to quantify the density of driver nodes needed for different graph configurations, providing a probabilistic understanding of controllability. The ability to incorporate randomness directly into the model aligns well with the stochastic nature of many real-world systems. Furthermore, the minimum matching model is computationally efficient, employing algorithms like Hopcroft-Karp to identify maximum matchings in bipartite graph representations. This scalability allows for analyzing large, complex random graphs, even with probabilistic edge formation. The insights gained from such an analysis can guide the design of networks to optimize controllability, such as by increasing connectivity to reduce the number of driver nodes. Using this model to measure the controllability of random graphs based on probability will provide a robust and mathematically grounded understanding of how network structures influence control dynamics in stochastic environments.
The Minimum Input Theory evaluates the controllability of directed networks by determining the minimum number of external inputs (\(N_D\)) required to control the system. This is achieved using the concept of a \textbf{maximum matching}, where unmatched nodes, called \textit{driver nodes}, must be directly controlled. The number of driver nodes is calculated as:
\[
N_D = N - |M^*|,
\]
where \(N\) is the total number of nodes and \(|M^*|\) is the size of the maximum matching. The fraction of driver nodes, defined as:
\begin{equation}
    n_D = \frac{N_D}{N},
    \label{eq:driver_node_res}
\end{equation}
serves as a metric for controllability, with lower \(n_D\) indicating easier control. Scale-free networks with heterogeneous degree distributions tend to have lower \(n_D\) due to hubs reducing unmatched nodes, whereas homogeneous networks require more driver nodes, making them less controllable. For scale-free networks with degree exponent \( \gamma \) in the large-\(\langle k \rangle\) limit, \( n_D \) is approximated as  
\begin{equation}
    n_D \approx \exp\left[-\frac{1}{2}\left(1 - \frac{1}{\gamma - 1}\right)\langle k \rangle\right],
    \label{eq:driver_node_degree}
\end{equation}
where \(\langle k \rangle\) is the average degree. This equation reveals that \(\gamma_c = 2\) acts as a critical exponent, beyond which the system can be controlled through a finite subset of nodes (\(n_D < 1\)), whereas for \(\gamma \leq \gamma_c\), all nodes must be individually controlled (\(n_D = 1\)). Denser networks, characterized by higher \(\langle k \rangle\), require significantly fewer driver nodes because the increased connectivity allows for greater redundancy and influence among nodes, thereby facilitating control through fewer input signals. In contrast, sparser networks are more challenging to control, requiring a larger fraction of driver nodes. Degree heterogeneity, defined as \( H = \Delta / \langle k \rangle \) with \(\Delta = \sum_{i,j}|k_i - k_j|P(k_i)P(k_j)\), further affects \( n_D \), with heterogeneous networks requiring more driver nodes due to disparities in node connectivity. Thus, while dense networks are easier to control, sparse and heterogeneous networks, such as those in biological systems or the Internet, require a significantly higher fraction of driver nodes, underscoring the difficulty of managing these complex systems.

\section{Numerical analyzing}
\label{sec:NumericalAnalyzing}
In this section, we delve into comprehensive numerical analyses to substantiate the theoretical constructs of our model. We begin by meticulously evaluating the criteria that define each fundamental network structure, providing illustrative examples that highlight their distinct characteristics and functionalities. Building upon this foundation, we proceed to analyze the intricate market relationships, meticulously assessing the associated risks of volatility and their implications for market stability. This involves modeling dynamic interactions and quantifying the impact of fluctuating market conditions on the network's performance. Finally, we investigate the controllability of the network, exploring how its structural properties influence our ability to manage and mitigate systemic risks effectively. Through this multi-faceted approach, we aim to demonstrate the robustness and applicability of our model in capturing the complexities of real-world economic networks.

\begin{figure}
    \centering
    \begin{minipage}[b]{0.21\textwidth}
        \centering
        \includegraphics[width=\textwidth]{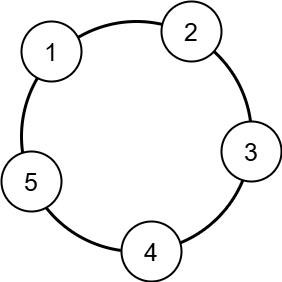}
        % \caption{Caption 1}
        \label{fig:cycle}
    \end{minipage}
    \hfill
    \begin{minipage}[b]{0.21\textwidth}
        \centering
        \includegraphics[width=\textwidth]{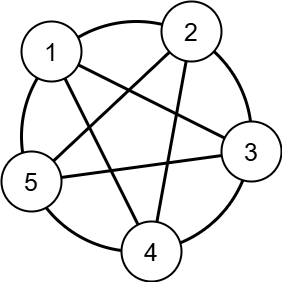}
        % \caption{Caption 2}
        \label{fig:complete}
    \end{minipage}
    \hfill
    \begin{minipage}[b]{0.21\textwidth}
        \centering
        \includegraphics[width=\textwidth]{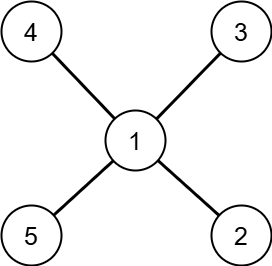}
        % \caption{Caption 3}
        \label{fig:star}
    \end{minipage}
    \hfill
    \begin{minipage}[b]{0.21\textwidth}
        \centering
        \includegraphics[width=\textwidth]{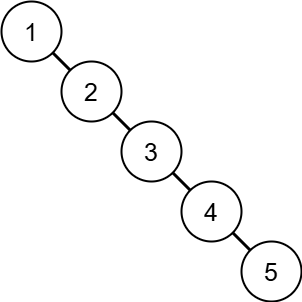}
        % \caption{Caption 4}
        \label{fig:path}
    \end{minipage}
    
    \caption{Consider three fundamental graph structures—(a) a cycle graph, (b) a complete graph, (c) a star graph and (d) a path graph. These graphs can represent basic models of a company's supply chain and relational structure: the cycle graph illustrating a closed, interdependent system, the path graph depicting a linear, sequential flow, and the star graph showing a centralized hub with connections radiating outward. All three graphs are symmetric, reflecting balanced, yet distinct, organizational frameworks.}
    \label{fig:four_basic_graph}
\end{figure}
Matrix \( A \), \( B \), and \( C \) are calculated for a cycle graph, and after one unit of time, taking into account an addition rate of 1 and a deletion rate of 0.5 per unit time. According to the model's definition, the probability for each node is defined in \eqref{eq:p_addition} and \eqref{eq:p_deletion}. The probabilities of addition (\( P_a \)) and deletion are determined as follows: \( P_a = \frac{2}{10} \) and \( P_d = n \times p_{d_i} = 0.5 \times \left(\frac{2}{10} \right) = \frac{1}{10} \), where \( n = 0.5 \) is the deletion rate. Matrix \( A \) considered:
\[
A = \begin{bmatrix}
0 & 0.4 & 0 & 0 & 0.4 \\
0.4 & 0 & 0.4 & 0 & 0 \\
0 & 0.4 & 0 & 0.4 & 0 \\
0 & 0 & 0.4 & 0 & 0.4 \\
0.4 & 0 & 0 & 0.4 & 0 
\end{bmatrix}
\]
The degree vector \( K \) is obtained as  \( K_i = \sum_{j}{S_{ij}} \):
\[
S = \begin{bmatrix}
0 & 1 & 0 & 0 & 1 \\
1 & 0 & 1 & 0 & 0 \\
0 & 1 & 0 & 1 & 0 \\
0 & 0 & 1 & 0 & 1 \\
1 & 0 & 0 & 1 & 0 
\end{bmatrix}
\]
To account for network dynamics, we construct the augmented matrix \( A_s \) by adding an additional row and column of zeros to \( A \):
\[
A_s = \begin{bmatrix}
A & \mathbf{0} \\
\mathbf{0}^\top & 0 
\end{bmatrix} = \begin{bmatrix}
0 & 0.4 & 0 & 0 & 0.4 & 0 \\
0.4 & 0 & 0.4 & 0 & 0 & 0 \\
0 & 0.4 & 0 & 0.4 & 0 & 0 \\
0 & 0 & 0.4 & 0 & 0.4 & 0 \\
0.4 & 0 & 0 & 0.4 & 0 & 0 \\
0 & 0 & 0 & 0 & 0 & 0 
\end{bmatrix}
\]
Matrix \( B \) is initialized as a zero matrix with the same dimensions as \( A_s \) and is updated to incorporate the addition probability \( P_a \), where \( b_{i6} = b_{6j} = 0.2 \) and \( b_{66} = 0 \). Therefore, \( B_{i,6} = B_{6,i} = 0.2 \times P_a = 0.2 \times \frac{1}{5} = 0.04 \), and \( B_{6,6} = 0 \). Thus, \( B \) becomes:
\[
B = \begin{bmatrix}
0 & 0 & 0 & 0 & 0 & 0.2 \\
0 & 0 & 0 & 0 & 0 & 0.2 \\
0 & 0 & 0 & 0 & 0 & 0.2 \\
0 & 0 & 0 & 0 & 0 & 0.2 \\
0 & 0 & 0 & 0 & 0 & 0.2 \\
0.2 & 0.2 & 0.2 & 0.2 & 0.2 & 0 
\end{bmatrix}
\]
Matrix \( C \) is derived by scaling the augmented matrix \( A_s \) with the deletion probability \( P_d \):
\[
C = P_d \times A_s = \begin{bmatrix}
    0 & 0.16 & 0 & 0 & 0.16 & 0 \\
    0.16 & 0 & 0.16 & 0 & 0 & 0 \\
    0 & 0.16 & 0 & 0.16 & 0 & 0 \\
    0 & 0 & 0.16 & 0 & 0.16 & 0 \\
    0.16 & 0 & 0 & 0.16 & 0 & 0 \\
    0 & 0 & 0 & 0 & 0 & 0 
    \end{bmatrix}
\]
Finally, to evaluate the network's controllability and stability, we compute the variances \( r_1 \) and \( r_2 \) as follows:
\[
L = (I - A)^{-1}
\]
\[
L_s = \left( I - (A_s + B - C) \right)^{-1}
\]
where \( I \) is the identity matrix corresponding to \( A \) and \( A_s \), respectively. These calculations facilitate the assessment of how addition and deletion dynamics influence the network's robustness and responsiveness to fluctuations.
Using equation \eqref{eq:aggregate_volatility}, we calculate the variance of volatility, specifically the values \( \text{var}\left(\sum_{j}{l_{ij}}\right) \) and \( \text{var}\left(\sum_{j}{l_{s_{ij}}}\right) \), for each matrix \( L \) and \( L_s \). The results are: the variance of volatility for \( L \) is 0 and for \( L_s \) is 0.588. Furthermore, the values computed for the \textbf{complete}, \textbf{star}, and \textbf{path} graphs, as shown in \hyperref[fig:four_basic_graph]{Figure~1}, are \( R_{\text{complete}} = (0, 0.588) \), \( R_{\text{star}} = (0.081, 0.096) \), and \( R_{\text{path}} = (0.346, 0.449) \). This corresponds to an 18.5\% change for the star and a 30\% change for the path. These calculations reveal how network dynamics vary across different topologies, with each graph exhibiting a distinct response to structural changes over unit time steps, as the complete and star networks adapt uniquely compared to the path graph.

The controllability of the star system is evaluated for a star graph topology, as detailed in \hyperref[app:appendix1]{Appendix~1}. The adjacency matrix \( \mathbf{A} \) is defined as a \( 6 \times 6 \) symmetric matrix, representing the star graph's structure, where a central node is connected to all other nodes, but there are no direct connections between the peripheral nodes. The input matrix \( \hat{\mathbf{E}} \) is parameterized to assign a unique control input (\( \epsilon_0 \)) to the central node and a common input (\( \epsilon_1 \)) to the peripheral nodes. Using the Kalman rank condition, the controllability matrix \( \mathcal{C}_{\hat{\mathbf{E}}} \), formed as \( \mathcal{C}_{\hat{\mathbf{E}}} = \begin{bmatrix} \hat{\mathbf{E}} & \mathbf{A}\hat{\mathbf{E}} & \mathbf{A}^2\hat{\mathbf{E}} & \cdots & \mathbf{A}^{n-1}\hat{\mathbf{E}} \end{bmatrix} \), is analyzed. Successive computations of \( \mathbf{A}^k \hat{\mathbf{E}} \) reflect the influence of the star graph topology on \( \mathcal{C}_{\hat{\mathbf{E}}} \). The rank of \( \mathcal{C}_{\hat{\mathbf{E}}} \), determined to be 2, indicates that the Kalman rank condition for full controllability is not satisfied. This analysis highlights the need for additional control inputs or structural modifications to achieve full controllability for the star graph configuration.
Based on the analysis of dynamic systems represented by graph structures, the controllability properties of path, cycle, and complete graphs were analyzed by constructing their adjacency matrices and evaluating the rank of their controllability matrices. For the \textbf{path graph}, full rank was achieved by the controllability matrix, indicating that the system is \textit{fully controllable} due to the linear structure and the direct propagation of control influence. Similarly, for the \textbf{cycle graph}, full controllability was observed when the input matrix \( \hat{\mathbf{E}} \) was diagonal with either distinct entries or uniform values (\( \epsilon, \dots, \epsilon \)), as the cyclic structure allows influence to propagate through the closed loop effectively. The \textbf{complete graph} was also found to be fully controllable, as its dense connectivity ensures direct influence between all nodes, making the system robust to the uniform or distinct control inputs provided by \( \hat{\mathbf{E}} \). In both cases, the rank of the controllability matrix was equal to the number of nodes, confirming full controllability.

The base network structure was analyzed, and subsequently, a near real-world network of industries based on the U.S. economic structure was interpreted. Based on the model and results presented in Section~\ref{sec:EvolvingProduction}, the probability distribution of degrees is described by Equation~\eqref{eq:evolveProb}, which accounts for the deletion and addition of nodes at rates \( n \) and \( m \), respectively. \cite{Atalay2011} estimated the model using real data and obtained specific values for each of these parameters\footnote{The dataset comprises 39,815 firm-year observations from the CRSP/Compustat database, focusing on publicly traded firms reporting major customers in accordance with SFAS 131 regulations. It includes 14,204 unique buyer-supplier relationships and additional firm attributes such as headquarters location, number of employees, total sales, and four-digit SIC industry classification.}. These parameters—vertex exit rate (\( q = 0.24 \)), average number of edges per vertex (\( \nu = 1.06 \)), fraction of edges connecting new vertices to previously existing firms (\( \delta = 0.75 \)), average growth rate of the number of vertices in the network (\( g = 0.04 \)), and estimated fraction of edges assigned uniformly across existing vertices (\( r = 0.18 \))—were utilized to calibrate the equation. Subsequently, for simplification, the probability distribution was estimated using a power-law distribution, and the exponent \( \gamma \) was approximated.

\begin{equation}
\label{eq:degree_distribution}
P(k) \propto k^{-\gamma}
\end{equation}

\begin{figure}[htbp]
    \centering
    \includegraphics[width=0.8\textwidth]{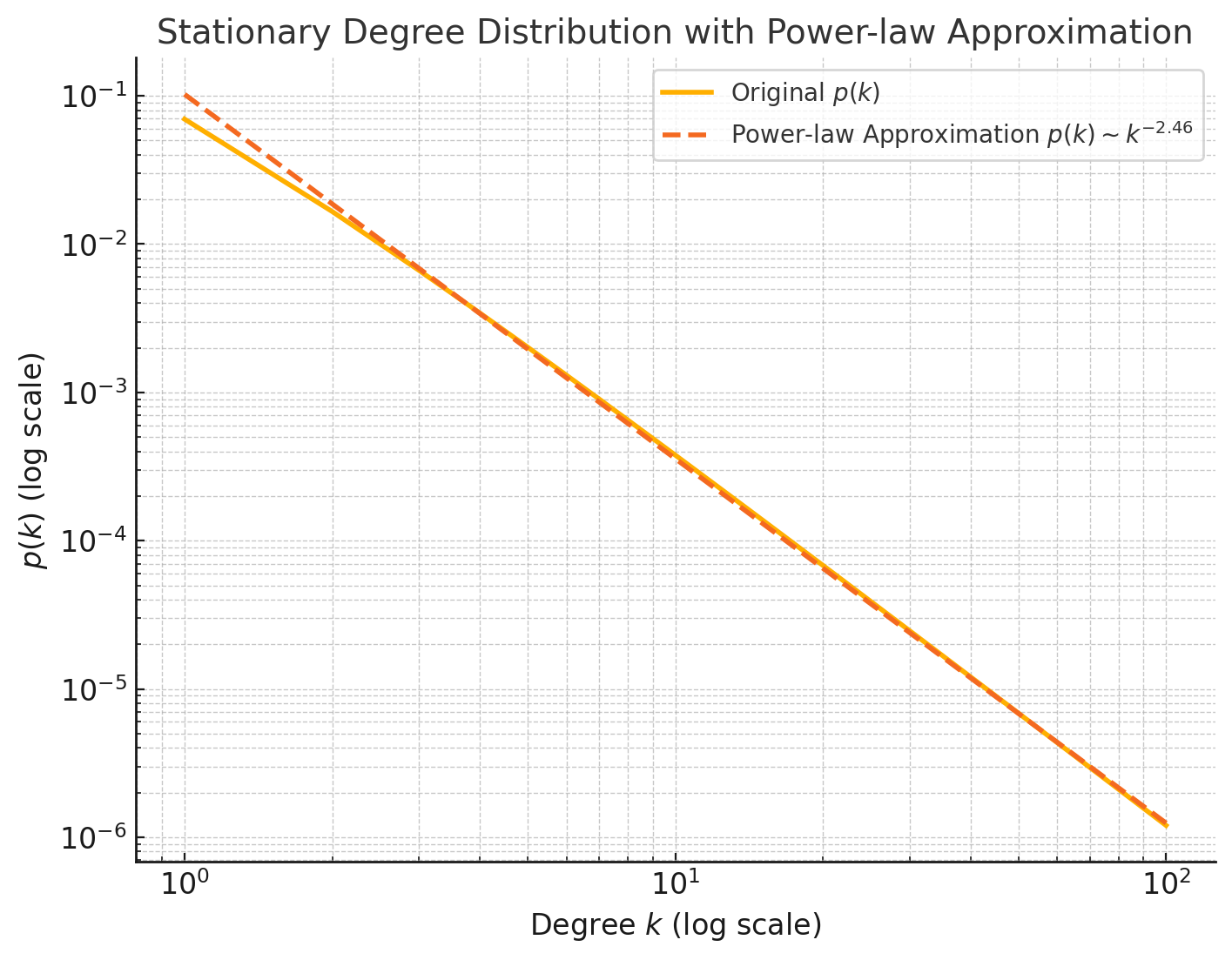} % Replace with your actual image filename
    \caption{Estimation of the Probability Distribution of Degrees. The power-law fit yields an exponent \(\gamma = -2.46\), indicating the scaling behavior of the network's degree distribution.}
    \label{fig:gamma_estimation}
\end{figure}

As depicted in Figure~\ref{fig:gamma_estimation}, the probability distribution of degrees follows a power-law distribution, confirming the theoretical model proposed. The estimated exponent \(\gamma = -2.46\) suggests a heavy-tailed distribution, which is characteristic of many real-world networks, including industrial networks. This value of \(\gamma\) aligns with previous studies, reinforcing the validity of the model and its applicability to the U.S. economic structure.

\begin{figure}[h!]
    \centering
    \begin{minipage}[b]{0.45\textwidth}
        \includegraphics[width=\textwidth]{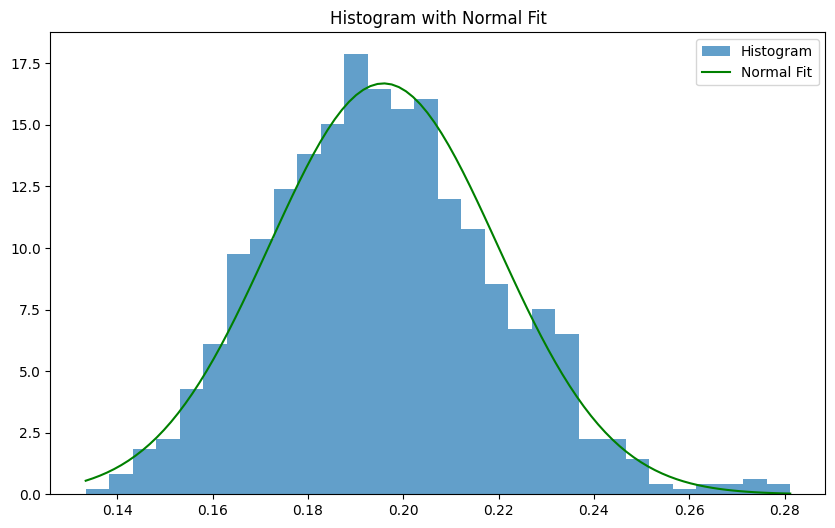}
        \caption{Histogram of Monte Carlo-simulated values with normal fit (\(\mu = 0.20, \sigma = 0.02\)).}
        \label{fig:histogram}
    \end{minipage}
    \hfill
    \begin{minipage}[b]{0.45\textwidth}
        \includegraphics[width=\textwidth]{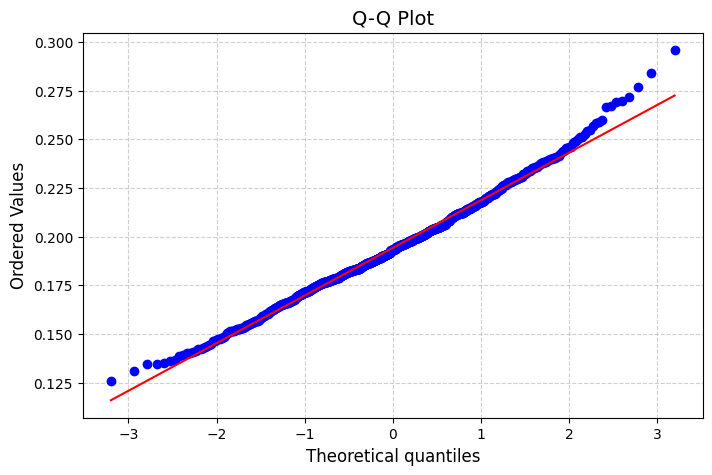}
        \caption{Q-Q plot comparing simulated values to the theoretical normal distribution.}
        \label{fig:qqplot}
    \end{minipage}
    % \caption{Visualizations of Monte Carlo simulations: (a) Histogram with normal fit and (b) Q-Q plot showing distribution alignment.}
    \label{fig:combined}
\end{figure}

The fitting process involves approximating the statistical properties of the simulated variable \( \text{var}(\nu_i) \) to a theoretical normal distribution. The histogram and Q-Q plot validate the quality of this fit, with the fitted parameters of mean (\(\mu\)) and standard deviation (\(\sigma\)) calculated as \(0.20\) and \(0.02\), respectively. The close alignment between the observed data and the theoretical normal distribution supports the reliability of the fitted model, which is crucial in ensuring accurate representation of the underlying data generation process. 

Furthermore, the tails of the distribution display slight deviations from normality, as evidenced by the Q-Q plot. For a normal fit with \(\mu = 0.20\) and \(\sigma = 0.02\), approximately \(4.6\%\) of the values are expected to fall outside \(\pm 2\sigma\), and \(0.3\%\) beyond \(\pm 3\sigma\). These tails are well-behaved, with no significant skewness or kurtosis, confirming the normal fit's adequacy for most data, though extreme tail events remain a measurable risk.

According to equation \eqref{eq:driver_node_res}, the fraction of driver nodes, \(n_D\), quantifies the proportion of nodes that must be directly controlled to achieve full controllability of the network. Based on our calculations, \(n_D \approx 0.550\). This means that at least \(55\%\) of the nodes in the network need to act as driver nodes to ensure complete control of the network dynamics.

The significance of this result lies in the fact that \(n_D\) is determined by the structural properties of the network, specifically the degree distribution exponent \(\gamma\) and the mean degree \(\langle k \rangle\). For a network with \(\gamma = 2.46\) and \(\langle k \rangle = 3.80\), the derived value of \(n_D\) highlights the critical threshold of driver nodes required to manage the network's behavior comprehensively. This underscores the dependency of network controllability on its topological features, as encapsulated by equation \eqref{eq:driver_node_degree}.

\section{Conclusion}
In this study, we have developed a novel theoretical framework to model the dynamic interdependencies within economic networks, emphasizing the probabilistic nature of firm interactions and their systemic implications. By integrating node deletion mechanisms—both individually and collectively—with a probability distribution, our model captures the inherent uncertainty of firm exits and their cascading effects on network stability. Leveraging numerical methods grounded in empirical data from the US economy, we derived a critical threshold for complete network controllability, demonstrating that approximately \(55\%\) of nodes must be actively managed to ensure structural and functional resilience. This finding not only aligns with prior research on degree distribution but also advances the discourse by quantifying controllability in complex economic systems.

Furthermore, the framework bridges controllability analysis with macroeconomic outcomes, enabling the calculation of production fluctuations as an expected value under probabilistic node addition and deletion. This dual capability—to model both control requirements and economic volatility—provides policymakers and stakeholders with a robust tool for anticipating systemic risks and designing interventions to mitigate disruptions. The model's flexibility in aggregating individual and collective firm dynamics underscores its applicability to diverse economic contexts, from crisis management to strategic industrial planning.

Future research could extend this work by incorporating empirical validations across different economies or exploring temporal adaptations to reflect real-time network evolution. Additionally, integrating heterogeneous firm-specific data (e.g., sectoral vulnerabilities, size-dependent probabilities) may refine the model's predictive accuracy. Ultimately, this framework contributes a foundational step toward understanding and governing the intricate web of modern economic networks, where controllability and adaptability are paramount in an era of escalating uncertainties.

% \appendix
% \section{Appendix}
% % this is Appenix A %
% Proof of equations
\appendix
\section*{Appendix 1}
\label{app:appendix1}
\subsection*{System Definition}

The adjacency matrix \( \mathbf{A} \) is parameterized as a \( 6 \times 6 \) matrix where each entry \( A_{ij} \) indicates the presence (\( a_{ij} = 1 \)) or absence (\( a_{ij} = 0 \)) of an edge between nodes \( i \) and \( j \). For an undirected star graph without self-loops, \( \mathbf{A} \) is symmetric with diagonal entries equal to zero:

\[
\mathbf{A} =
\begin{bmatrix}
0 & a_{12} & a_{13} & a_{14} & a_{15} & a_{16} \\
a_{21} & 0 & 0 & 0 & 0 & 0 \\
a_{31} & 0 & 0 & 0 & 0 & 0 \\
a_{41} & 0 & 0 & 0 & 0 & 0 \\
a_{51} & 0 & 0 & 0 & 0 & 0 \\
a_{61} & 0 & 0 & 0 & 0 & 0
\end{bmatrix}.
\]

\subsubsection*{Step 1: Define \( \hat{\mathbf{E}} \)}

The input matrix \( \hat{\mathbf{E}} \) is parameterized with \( \epsilon_0 \) as the control input for the first node and \( \epsilon_1 \) for all other nodes:

\[
\hat{\mathbf{E}} =
\begin{bmatrix}
\epsilon_0 & 0 & 0 & 0 & 0 & 0 \\
0 & \epsilon_1 & 0 & 0 & 0 & 0 \\
0 & 0 & \epsilon_1 & 0 & 0 & 0 \\
0 & 0 & 0 & \epsilon_1 & 0 & 0 \\
0 & 0 & 0 & 0 & \epsilon_1 & 0 \\
0 & 0 & 0 & 0 & 0 & \epsilon_1
\end{bmatrix}.
\]

This parameterization allows for differentiated control inputs across the nodes, with the first node having a unique input and the remaining nodes sharing a common control parameter.

\subsubsection*{Step 2: Compute \( \mathbf{A}\hat{\mathbf{E}} \)}

Multiplying the adjacency matrix \( \mathbf{A} \) with \( \hat{\mathbf{E}} \) results in:

\[
\mathbf{A}\hat{\mathbf{E}} =
\begin{bmatrix}
0 & \epsilon_1 & \epsilon_1 & \epsilon_1 & \epsilon_1 & \epsilon_1 \\
\epsilon_0 & 0 & 0 & 0 & 0 & 0 \\
\epsilon_0 & 0 & 0 & 0 & 0 & 0 \\
\epsilon_0 & 0 & 0 & 0 & 0 & 0 \\
\epsilon_0 & 0 & 0 & 0 & 0 & 0 \\
\epsilon_0 & 0 & 0 & 0 & 0 & 0
\end{bmatrix}.
\]

\subsubsection*{Step 3: Compute Higher Powers of \( \mathbf{A} \)}

Higher powers of \( \mathbf{A} \) multiplied by \( \hat{\mathbf{E}} \) show the following patterns:

\[
\mathbf{A}^2\hat{\mathbf{E}} =
\begin{bmatrix}
5\epsilon_0 & 0 & 0 & 0 & 0 & 0 \\
0 & \epsilon_1 & \epsilon_1 & \epsilon_1 & \epsilon_1 & \epsilon_1 \\
0 & \epsilon_1 & \epsilon_1 & \epsilon_1 & \epsilon_1 & \epsilon_1 \\
0 & \epsilon_1 & \epsilon_1 & \epsilon_1 & \epsilon_1 & \epsilon_1 \\
0 & \epsilon_1 & \epsilon_1 & \epsilon_1 & \epsilon_1 & \epsilon_1 \\
0 & \epsilon_1 & \epsilon_1 & \epsilon_1 & \epsilon_1 & \epsilon_1
\end{bmatrix},
\]
\[
\mathbf{A}^3\hat{\mathbf{E}} =
\begin{bmatrix}
0 & 5\epsilon_0 & 5\epsilon_0 & 5\epsilon_0 & 5\epsilon_0 & 5\epsilon_0 \\
5\epsilon_0 & 0 & 0 & 0 & 0 & 0 \\
5\epsilon_0 & 0 & 0 & 0 & 0 & 0 \\
5\epsilon_0 & 0 & 0 & 0 & 0 & 0 \\
5\epsilon_0 & 0 & 0 & 0 & 0 & 0 \\
5\epsilon_0 & 0 & 0 & 0 & 0 & 0
\end{bmatrix}.
\]

\subsection*{Final Controllability Matrix with Modified \( \hat{\mathbf{E}} \)}

The controllability matrix \( \mathcal{C}_{\hat{\mathbf{E}}} \) is parameterized as:

\[
\mathcal{C}_{\hat{\mathbf{E}}} = \begin{bmatrix}
\hat{\mathbf{E}} & \mathbf{A}\hat{\mathbf{E}} & \mathbf{A}^2\hat{\mathbf{E}} & \mathbf{A}^3\hat{\mathbf{E}} & \cdots & \mathbf{A}^5\hat{\mathbf{E}}
\end{bmatrix}.
\]

Substituting the computed matrices:

\[
\mathcal{C}_{\hat{\mathbf{E}}} =
\begin{bmatrix}
\epsilon_0 & 0 & 5\epsilon_0 & 0 & \cdots & \ast \\
0 & \epsilon_1 & 0 & 5\epsilon_0 & \cdots & \ast \\
0 & \epsilon_1 & 0 & 5\epsilon_0 & \cdots & \ast \\
0 & \epsilon_1 & 0 & 5\epsilon_0 & \cdots & \ast \\
0 & \epsilon_1 & 0 & 5\epsilon_0 & \cdots & \ast \\
0 & \epsilon_1 & 0 & 5\epsilon_0 & \cdots & \ast
\end{bmatrix}.
\]

\subsection*{Rank of \( \mathcal{C}_{\hat{\mathbf{E}}} \)}

The rank of the controllability matrix \( \mathcal{C}_{\hat{\mathbf{E}}} \) is calculated as:

\[
\text{rank}(\mathcal{C}_{\hat{\mathbf{E}}}) = 2.
\]

The modified diagonal input matrix \( \hat{\mathbf{E}} \), with \( \epsilon_0 \) for the first node and \( \epsilon_1 \) for the remaining nodes, results in a controllability matrix of rank 2. For a system with \( n = 6 \), this indicates the system is not fully controllable. Achieving full controllability would require additional independent inputs or changes to the graph structure.

\medskip

% for citation we should make all cites correct
\bibliography{reference}

\end{document}